\documentclass[]{mn2e}

\usepackage{amssymb}
\usepackage{graphicx}
\usepackage{amsmath}
\usepackage{graphics}

\usepackage{color}

\title[Phonons in molecules] {Molecular phonons and their absorption/emission spectra from the far IR to microwaves}
\author[R. Papoular]{R. Papoular$^{1}$\thanks{E-mail:
papoular@wanadoo.fr}\\
$^{1}$Service d'Astrophysique and Service de Chimie Moleculaire,\\
CEA Saclay, 91191 Gif-s-Yvette, France}

\begin{document}

   \maketitle
\label{firstpage}

\begin{abstract}
Together with their fingerprint modes, molecules carry coherent vibrations of all their atoms (phonons). Phonon spectra extend from $\sim$20 to more than $10^{4}\,\mu$m, depending on molecular size. These spectra are discrete but large assemblies of molecules of the same family, differing only by minor structural details, will produce continua. As such assemblies are expected to exist in regions where dust accumulates, they are bound to contribute to the observed continua underlying the Unidentified Infrared Bands and the 21-$\mu$m band of planetary nebulae as well as to the diffuse galactic emission surveyed by the \emph{Planck} astronomical satellite and other means. The purpose of this work is to determine, for carbon-rich molecules, the intensity of such continua and their extent into the millimetric range, and to evaluate their detectability in this range. The rules governing the spectral distributions of phonons are derived and shown to differ from those which obtain in the solid state. Their application allow the extinction cross-section per H atom, and its maximum wavelength, to be determined as a function of molecular size and dimensionality. Chemical modeling of more than 15 large molecules illustrate these results. It is found that the maximum phonon wavelength  of a 2D structure increases roughly as the square of  its larger dimension. Spectral energy distributions were computed as far as 4000 $\mu$m, for molecules up to 50 \AA{\ } in length.

\end{abstract}


\section{Introduction}
Phonons in extended media have long been the subject of detailed investigations, in particular of their DOM (density of modes, $g$) and velocity, $V$, as a function of angular frequency $\omega$ (e.g. see Kittel 2005). Graphite, in particular, was abundantly investigated  from the early work of Nicklow, Wakabayashi and Smith \cite{nic} to the recent one by Mohr et al. \cite{moh}. By contrast, the study of vibrations in finite structures like molecules, has largely been limited to their fingerprints, i.e. those vibrations which involve only a handful of closely bonded atoms within the structure (functional groups) and usually fall above 500 cm$^{-1}$ (e.g. see Colthup et al. 1990). Molecular phonons differ from fingerprints in that they involve the whole structure; their spectra are quasi-periodic and extend from $\sim$20 to more than $10^{4}\,\mu$m (depending on molecular size) and so may contribute to the continuum that is observed to underly the UIB spectra of galaxies and planetary nebulae:  see, for instance, Smith et al. \cite{smi}, Compi\`egne et al. \cite{com}, Zhang and Kwok \cite{zha}, Fischer et al. \cite{fis}.

Low-frequency phonons are presently of particular interest to astrophysics as electromagnetic emission in the microwave/millimeter ranges has now been detected, measured and surveyed all over the sky  (e.g. see Finkbeiner et al. \cite{fin}; Ghosh et al. (\cite{gho} and bibliography therein). More recently, Ade et al. \cite{ade14} and Abergel et al. \cite{abe14}, analyzing data from the \emph{Planck} astronomical satellite, obtained quasi-continuous spectra from 20 to 3000 GHz (100 to 15000 $\mu$m). In this range, several emission sources appear to be in action: thermal dust, spinning and magnetic dust (Draine and Lazarian (1998), Lazarian and Draine (2000)), free-free electrons, synchrotron. Much effort has gone into combining these various sources so as to fit observations. The purpose, here, is more limited and focused on molecular vibrations that may be excited either thermally or otherwise. We seek to define the physical (structure, size) and chemical (composition) conditions to be met by a molecule for its low-frequency vibrations to contribute significantly to the observed spectra, and to determine the nature of the vibrations involved. This also differs from Papoular \cite{pap14a} and \cite{pap14b}, in which the maximum wavelength was 1000 $\mu$m and  only a mixture of molecules was considered but not their  respective contributions to the emission.  Also, this work can be considered a first step in extrapolating the extinction curve of carbonaceous dust into the microwave range.

Fortunately, the study of molecular phonons is now made easier by computer modeling of finite, but very large, bonded structures (hundreds of atoms). In the course of such an investigation, it turned out that the frequency dependences of $g$ and $V$, for the transverse acoustic phonons of 1D and 2D molecules, differed from the well-known Debye laws for the case of bulk media (see Kittel 2005: $g\propto\omega^{2}$ and $V$ constant, in the limit of low frequencies, in which we are primarily interested here). While 2D molecules could be likened superficially to very thin plates, the theoretical treatment of the latter involves Young's modulus and Poisson's ratio (see Reddy 2007) so it cannot be extended to molecules.

Based on the study of the transverse and longitudinal acoustic phonons (TA and LA, respectively) of simple linear structures, it is shown here that the reason for this behavior of untethered molecules is that the vibrational mechanics is dominated by the force which resists bond bending, so the sound speed increases with frequency, by contrast with the constant sound speed in crystals (Sec. 2). 
 
 In most cases, the topological complexity of the structure does not allow direct analysis of motion for each individual phonons; only the DOM is readily available from modeling. But what is needed to predict the lowest natural frequency of a given structure as a function of its size, will turn out to be the sound velocity. A mathematical analysis of the general case led us to a simple formula by means of which $V$ can be derived from the sole knowledge of $g$. At low frequencies, $g$ and $V$ are governed by power laws; another outcome of our analysis is a set of relations between the dimensionality of the structure, the type of forces in action and the indexes of the two power laws which govern $g$ and $V$, respectively (Sec. 3). Note that the mentioned dimensionality refers here to the carbonaceous backbone of a molecule.
 
 In Sec. 4, the theory developed in Sec. 3 is applied to several 2D molecules which are shown to obey the predicted relations even though they may be bent and/or warped. The lowest phonon  frequency (always that of a TA phonon) is shown to scale like the square of the major dimension of the molecule. In Sec. 5, the theory is further illustrated with several individual molecules.

In Sec. 6, the integrated band intensity (absorption) spectrum is computed for several large molecules. The corresponding extinction cross-section per H atom in the ISM (InterStellar Medium) is computed. The absorption spectrum is used in Sec. 7 to obtain the emission spectrum of a particle in thermal equilibrium with its environment. The case where the particle is too small to reach thermal equilibrium is then considered, detailing a procedure used to obtain the corresponding emission spectrum, which is distinctly different than in the previous case. It is concluded that reasonable amounts of C atoms locked in similar structures are sufficient to emit detectable radiation as far as 4000 \AA{\ } at least.

\section{Phonons in linear atomic chains}
 
 In the course of this work, a large number of carbonaceous molecules were simulated by means of various algorithms of computational organic chemistry as embodied in the Hyperchem software provided by Hypercube, Inc., and described in detail in their publication HC50-00-03-00. The semi-empirical (mixed quantum/classical) PM3 method, initially developed by J. P. Stewart \cite{ste}, was used throughout, as its algorithms are specially optimized for  carbonaceous structures; its accuracy for vibrational frequencies is a few percent. It does not require the initial selection and setting of particular force fields.

The procedure for each molecule is as follows. First, the molecule is schematically represented on the computer screen. Then, the code is prompted to optimize the molecular configuration by adjusting the various bond lengths and angles so as to minimize the total bond energy. At that point, a Normal Mode analysis can be automatically performed, which delivers the wave number and integrated band intensity (oscillator strength) of each vibrational mode of the system.

As an example, consider a linear chain of 20 doubly-bonded carbon atoms, =C=. In order to impose a strict dimensionality of 1, it is terminated at each end with $\equiv$C-H; this is an ansatz provided by the simulation code under ``Allow arbitrary valence'' or ``Allow ions'', and boils down to adding a local electric 
charge.  Two alternative terminations were also used: CH2 and C=O, and gave similar results.

 The atomic motions were monitored on the screen for each normal mode found by the code. Three types of motion were found: a) transverse atomic displacements parallel to the screen plane; b) transverse displacements orthogonal to that plane; c) longitudinal displacements parallel to the central line. The first two types have essentially identical frequencies, the third has distinct frequencies shifted higher up. In all 3 cases, the amplitudes of atomic excursions display, all along the chain, a clear stationary sinusoidal pattern of standing waves which can be associated with harmonic waves traveling in opposite directions, whose wavelength along the chain
 decreases with increasing frequency. At each transition from one normal mode to the next, the number of half sinusoids abruptly increases by one. This is in line with the classical analysis of regular crystal lattices (see Kittel 2005). It is therefore possible to introduce here, too, the notions of phonon and wave vector  $k=2\pi/\lambda_{p}$, where $\lambda_{p}$ is the wavelength of the phonon at the corresponding  angular frequency,  $\omega=2\pi\,\nu\,\mathrm{c}$  where  
 $\nu$ is the  frequency  (wavenumber) and c is the velocity of light in vacuum; the density of modes, $g(\omega)$, like the density of states (DOS), can also be defined as the number of normal modes per unit frequency. By definition, the phase velocity is then $V=d\omega/dk$ . If the normal modes are labeled by integer $i$, starting from 1, in order of increasing frequency, then $\lambda_{p}(i)=2L_{\mathrm{mol}}/i$, where $L_{\mathrm{mol}}$ is the length of the chain; hence  $k(i)=i\pi/L_{\mathrm{mol}}$.

 A careful monitoring of the vibrational patterns shows that the expression for $\lambda_{p}(i)$ must be divided by a factor $f$, which depends slightly on the molecular length and terminations, but remains in the range of 1.5 to 2 (here, 1.73). This is due to the fact that, in the present cases, the boundary conditions are not forced upon the molecule by stretching it between two fixed points, but are determined by the electrical properties of the H termination at each end of the chain; as
 a result, the chain ends do not coincide with nodes of the standing wave.

\subsection{Transverse vibrations}

\begin{figure}
\resizebox{\hsize}{!}{\includegraphics{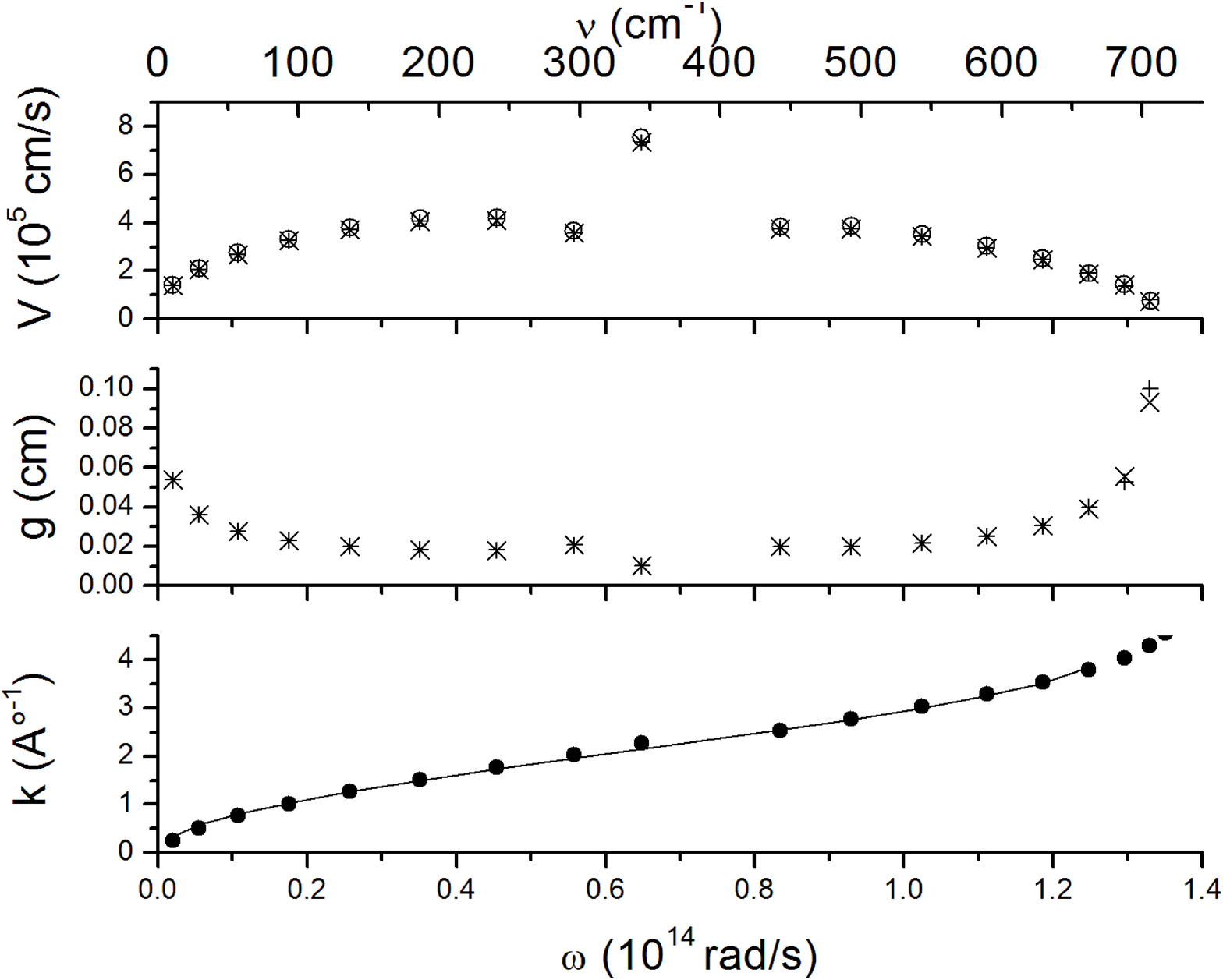}}
\caption[]{Transverse phonons. From bottom up: A) wave vector $k(\nu)$; dots: from chemical modeling; line: fit with eq. 4. B) DOM $g(\nu)$: crosses and pluses for parallel and perpendicular vibrations, respectively. C) $V(\nu)$: crosses and pluses for parallel and perpendicular vibrations, respectively; circles: fit with eq. 12. All from the modeling of the chain of 20 doubly-bonded C atoms.}
\label{Fig:transv15}
\end{figure}

 Figure \ref{Fig:transv15} shows, as a function of the frequency, $\nu$: A) the wave vector, $k$; B)the DOM, $g$; C) the sound velocity, $V=d\omega/dk$, of the transverse vibrations, for atomic displacements parallel ($\times$) and perpendicular (+) to the screen plane, all from the modeling of the chain of 20 doubly-bonded C atoms. As expected, there is a near coincidence of the 2 sets of points in each of graphs B and C . These vibrations, of course, are reminiscent of plate vibrations, and correspond to the transverse acoustic (TA) waves of solid state physics.

Note the sound speed variation at low frequencies. This trend is confirmed by the modeling of  six other linear doubly-bonded carbon chains, made of 12, 20, 40, 80, 120 and 160 C atoms, and terminated with CH2 at each end. The same procedure as above yields the TA sound velocities drawn in Fig. 2, together with that of  Fig. \ref{Fig:transv15} (for the sake of clarity, only one of the two sets of transverse vibrations is included). To the accuracy of the procedure, all 7 sound velocities are nearly coincident, except for the chain of Fig. \ref{Fig:transv15}, which is only slightly different, this being due to the different terminations. As expected, $V$ is nearly independent of the chain length. The logarithmic scales highlight the power-law dependence of the sound speed on frequency at low frequencies, $V=3.2\,10^{4}\,\nu^{0.5}$.

 \begin{figure}
\resizebox{\hsize}{!}{\includegraphics{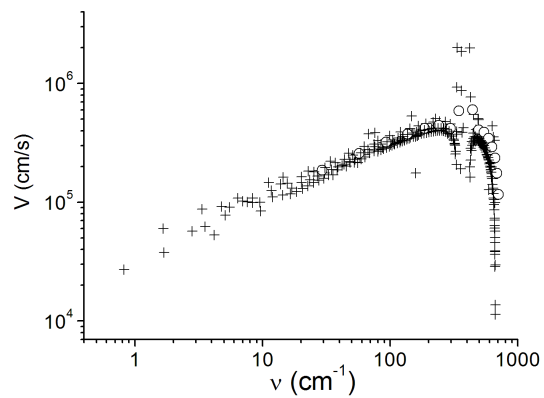}}
\caption[]{Same as graph C in Fig. \ref{Fig:transv15} for 7 linear chains of different lengths, including that of  Fig. \ref{Fig:transv15} (circles), except here, for clarity, only one of the two sets of the  transverse vibrations is represented. Note the logarithmic scales, the wave numbers in abscissa and the gap near 440 cm$^{-1}$. The rising part of the graph follows roughly the law $V=3.2\,10^{4}\,\nu^{0.5}$.}
\label{Fig:nuV7}
\end{figure}

Returning to Fig. \ref{Fig:transv15}, note that the DOM goes to infinity at $\nu=0$ and $\simeq700$ cm$^{-1}$; correlatively, by the definitions of $V$ and $g$, $V$ goes to 0. These are the only singularities predicted by Van Hove \cite{van} for infinite linear chains. With finite chains, however, another type of singularity occurs near the minimum of the DOM at $\omega_{}/2$, characterized by a gap in mode frequencies, wider or narrower depending on the type of terminations. This is due to a chance resonance between one TA mode and any other strong mode of the system, which gives rise to a strong mixed mode. The latter then perturbs the distribution of neighboring modes by pushing them aside. This in turn, locally disturbs the DOM and sound speed, all the more where the DOM is low, as is apparent in Fig. \ref{Fig:transv15}. In the present case, this occurs for the second longitudinal mode at 439 cm$^{-1}$, which resonates with transverse modes at 442.6 and 443 cm$^{-1}$, as shown by the following study of longitudinal acoustic (LA) branch of vibrations.

\subsection{Longitudinal vibrations}
\begin{figure}
\resizebox{\hsize}{!}{\includegraphics{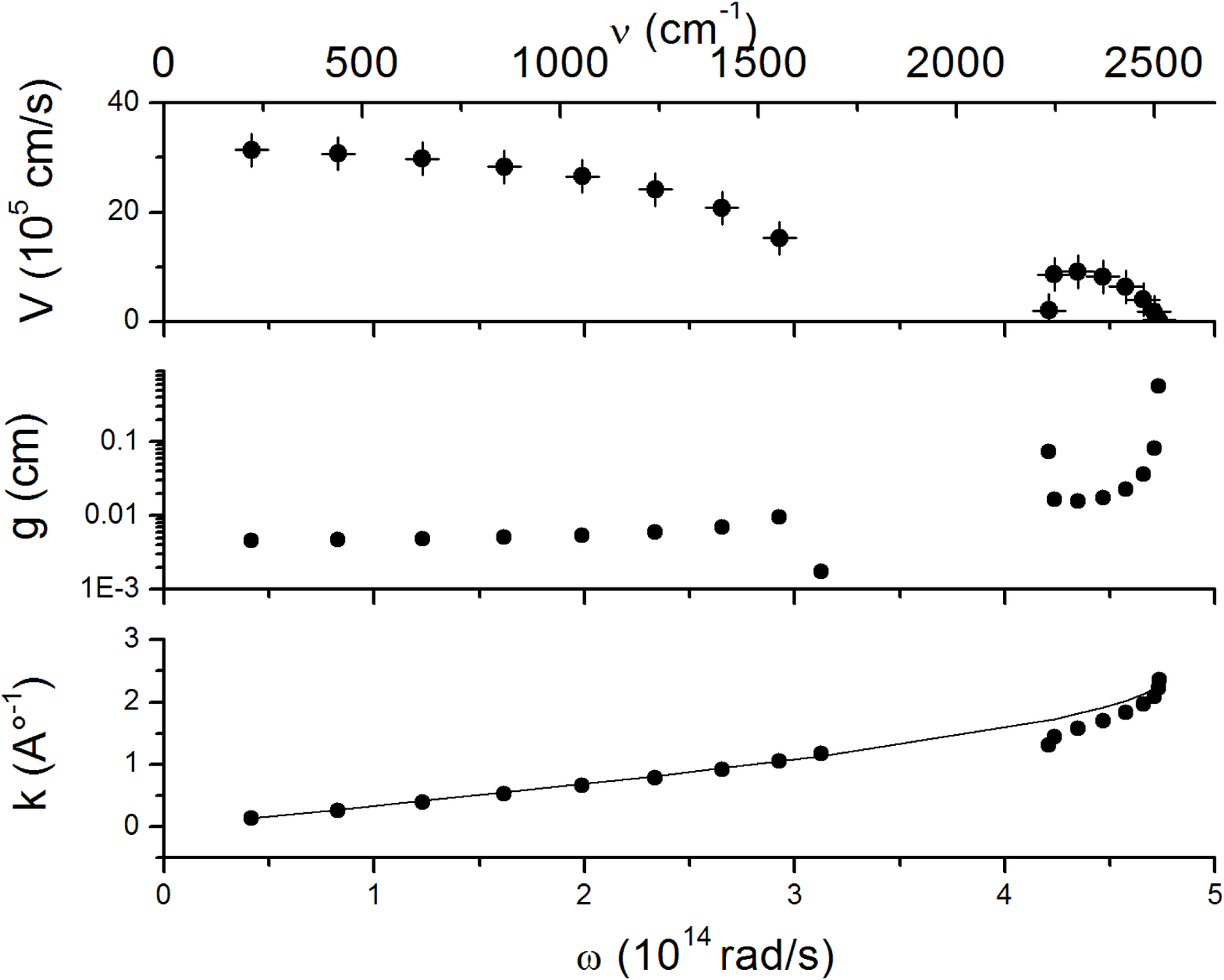}}
\caption[]{Longitudinal phonons. From bottom up: A)wave vector $k(\omega)$; dots: from chemical modeling; line: fit with eq. 6. B)DOM $g(\omega)$. C)sound speed $V(\omega)$; dots: chemical model; +: fit with eq. 12.}
\label{Fig:longit15}
\end{figure}

In monitoring, on the computer screen, atomic displacements
 relative to the chain backbone, in each normal mode, it is straightforward to distinguish longitudinal from transversal vibrations.  Fig. \ref{Fig:longit15} shows the behavior of wave vector, DOM and sound velocity for the longitudinal vibrations of the same structure: by contrast with the case of transverse vibrations, the latter two are nearly constant at low frequencies. Note that, in the familiar case of 3D media, while $V$ behaves likewise, $g(\nu)$, for its part, scales like $\nu^{2}$, instead.
Remarkably, the sound speed of longitudinal waves is an order of magnitude higher than that of the transversal ones. As a consequence, the first longitudinal mode occurs at a much higher frequency than its transverse counterpart. The second is the one that causes the frequency gap in Fig. \ref{Fig:transv15}, as explained above. There is also a gap in the longitudinal spectrum, but again at a much higher frequency, 855 cm$^{-1}$. In this case, it is due to interaction with a strong CH bending mode, which is clearly revealed on the modeling screen, at the chain end.

In this work, modes involving strong motions of terminal heteroatoms, and extending only slightly into the chain, were observed on all structures. They occur at the high frequency end of the transverse phonon spectral range, where end modes are strongly coupled to the last transverse mode. They are reminiscent of the ``surface'' modes studied by Wallis \cite{wal}, who showed that they can affect the thermal properties of very small powder grains. Here, however, these modes are relatively very few, so they were not given particular attention, especially considering that we are primarily interested in larger grains and lower frequencies. 

Also note the absence of an optical phonon branch, as the atoms on the chain are identical.

\subsection{Vibrational mechanics of phonons}
Now, why should the velocity of transverse waves depend on the wave number, as documented above? A plausible surmise is that the main stress acting on the transverse vibrations considered above is not the internal longitudinal tension (since the string is not stretched) but that which is induced by the bending of adjacent atomic bonds. It is known, indeed, that, for each atom type, there is a preferred angular distribution of the valence electrons. In the present case, the double bonds on both sides of an atom resist any stress that tends to bend their alignment. Molecular mechanics models (CHARMM, Gaussian, etc.) account for this by introducing explicitly potential term in $\theta^{2}$, where $\theta$ is the bending angle. In semi-empirical modeling codes, like PM3 here, the effect automatically results from the quantal description of the valence electrons. In the present instance, by noting the increase, $\Delta\,U$, in potential energy upon bending one of the pairs of bonds step by step from 0 to 5 deg, it was found that $\Delta\,U=\epsilon\theta^{2}$, where $\epsilon=1.85\,10^{-12}$ erg/rad$^{2}$  (8$\,10^{-3}$ kcal/mol/deg$^{2}$) and $\theta$ is in radians.

Let us assume that this is the only potential at work, and postulate for $\theta$ a trial solution in the form of a traveling wave, expi($\omega t+rka$), where $r$ is an integer designating the rank of an atom in the chain, and $a$ the distance between adjacent atoms, 1.28 \AA{\ }. Also, let $y_{r}$ be the transverse displacement of the $r$th carbon atom along the chain. Then, the angular displacement of the adjacent C=C bond relative to the chain's backbone is 
\begin{equation}
\alpha_{r}\simeq\mathrm{sin}\alpha=\frac{2y_{r}-y_{r-1}-y_{r+1}}{2a}=\mathrm{sin}^{2}(ka/2)\frac{2y_{r}}{a},
\end{equation}
and $\theta=2\alpha$. The total potential and kinetic energies of the system are, respectively,

\begin{equation}
U=4\epsilon\sum_{r}\alpha_{r}^{2}\,\, \mathrm{and}\,\, T=\sum_{r}\frac{m\dot y_{r}^{2}}{2}.
\end{equation}
where $m$ is the atomic mass, and it was assumed that $U$ is not dependent on the $\dot\alpha_{r}$'s. One can then write an independent Lagrange's equation of the form
\begin{equation}
\frac{d}{dt}\frac{\delta\,T}{\delta\,\dot\alpha_{r}}-\frac{\delta\,T}{\delta\,\alpha_{r}}+\frac{\delta\,U}{\delta\,\alpha_{r}}=0
\end{equation}
for each C atom in the chain, which gives the condition of existence of the trial solution:
 \begin{equation}
ma^{2}\omega^{2}=32\,\epsilon\,\mathrm{sin}^{4}(\frac{ka}{2}).
\end{equation}
  The best fit of this to $k(\omega)$ in Fig. \ref{Fig:transv15} indeed requires $\epsilon=1.85\,10^{-12}$  erg/rad$^{2}$ and is drawn as a line superimposed on the dots. The maximum angular frequency obtains for $ka/2=\pi/2$ (i.e. when the phonon wavelength reaches its minimum, $2a$) and is then 
 \begin{equation}
 \omega_{\mathrm{max}}=(\frac{32\,\epsilon}{ma^{2}})^{0.5}=1.31\,10^{14}\,\mathrm{rad/s}. 
\end{equation}
The inflexion point $\omega_{c}=\omega_{}/2$ is at $0.66\,10^{14}$ rad/s, where $g$ is minimum and $V$ maximum, as confirmed by  Fig. \ref{Fig:transv15}. The interpretation of the modeling results for transverse vibrations is therefore validated. The physical reason for the variation of $V$ is that the degree of bending of the molecule depends on $k$ and, therefore, on $\omega$.

 The same procedure can be applied to the longitudinal vibrations shown in  Fig. \ref{Fig:longit15}. Here the stretching potential involved is quadratic in the variable $a$:
  $\Delta\,U=\beta\Delta\,a^{2}$. However, the fit of $k(\omega)$ now requires
 \begin{equation}
m\omega^{2}=4\beta\,\mathrm{sin}^{2}(\frac{ka}{2}),
\end{equation}
as in the classical treatment of solids, with $\beta=1.12\,10^{6}$ erg/cm$^{2}$ (617 kcal/mol/\AA{\ }$^{2}$)
 and $\omega_{\mathrm{max}}=4.8\,10^{14}$ rad/s in the present instance. While the near constancy of $V$ at low frequencies naturally derives from this equation, the near constancy of $g$ is intriguing. The reason for it will become clear below.
 
 Note that, in both the transverse and longitudinal cases, the power index of $k$ is even, indicating the existence of two waves traveling in opposite directions for each resonant frequency.

In dealing with more complex molecular structures, it is no longer possible, by simple monitoring of each vibration on the computer screen, to distinguish phonons belonging to the same family, determine their wave vector, $k$, and apply the same procedure. The sole quantity that can still be clearly defined and computed is the DOM. This difficulty can be circumvented as described next.

\section{General relations for n-dimensional molecules}
 Consider first a periodic structure in $n$ dimensions, with size $L$ in every dimension. Let $k_{i}$ be the wave vector of vibration $i$. According to the Born-von Karman (cyclic) boundary conditions, the structure can only sustain waves such that its length is an integral number of half-wavelengths. Then, the number of possible vibrations, or states of the system, with $k\leq\,k_{N}$ is 
 
\begin{equation}
N=(\frac{2k_{N}}{2\,\pi/L})^{n},
\end{equation}
where the factor 2 in the numerator accounts for the fact that $k$ can be positive as well as negative. Then, the density of modes around $k_{N}$ (in $k$-space) is
\begin{equation}
w(k_{N})=\frac{dN}{dk_{N}}=n\frac{2^{n}k_{N}^{n-1}}{(2\,\pi/L)^{n}},
\end{equation}
and the DOM is 
\begin{equation}
g(\omega)=\frac{w(k_{N})}{d\omega/dk}.
\end{equation}
 Hence, 
 \begin{equation}
g(\omega)d\omega=Ck^{n-1}dk\,;\, C\equiv\,n(\frac{L}{\pi})^{n}.
\end{equation}
Let $G(\omega)=\int g(\omega)d\omega$, which is also the phonon rank when the phonons are sorted in order of increasing frequency. Then
\begin{equation}
k(\omega)=(\frac{nG(\omega)}{C})^{1/n}
\end{equation}
and
\begin{equation}
V(\omega)=\frac{d\omega}{dk}=\frac{nL}{\pi}\frac{G^{1-1/n}}{g}.
\end{equation}
  Thus, $k$ and, more importantly, $V$, can be derived directly from the DOM, which, for large, complex molecules, is the only readily computable quantity. Note that $g$ and $G$ are proportional to the total number of modes $N$, which is known to be nearly equal to 3 times the number of atoms (for large molecules). This, in turn is proportional to the ``volume'' of the molecule (for molecules of the same structure), i.e. the product of all the dimensions,$L_{1}...L_{n}$. Therefore, in the general case where all dimensions are not equal in size, $L$ in eq. 12 should be equated with the geometric mean of the dimensions,  $(L_{1}...L_{n})^{\frac{1}{n}}$. In any case, the dependencies of $g$ and $G$ on $L$ cancel out the factor $L$ in Eq. 12, so the phonon velocity is in fact independent of size, as it should.
 On the other hand, the longest wavelength sustained by the structure is not determined by $L$ but by the longest dimension of the structure.

 This treatment is only valid in spectral ranges where no overlapping of vibrational modes of different types (transverse/longitudinal) occur, which is only the case for the lowest-frequency transverse phonons. In that case, the results of the theoretical treatment for the speed of transverse waves agree with those of chemical modeling to better than a factor 2. This difference itself can be traced back to the absence of the correcting factor $f$ (Sec. 2.1) in the calculations, because of the assumed Born-Karman conditions.
 
 Now, since we are only interested in the lowest frequencies, we are entitled to set, as suggested by the log-log graph in Fig. 2, and as an approximation, 
\begin{equation}
g(\omega)\propto \omega^{m};\,\, V(\omega)\propto\omega^{p}.
\end{equation}
It follows that
\begin{equation}
V(\omega)\propto\omega^{1-1/n-m/n},
\end{equation}
and, finally,
\begin{equation}
p+\frac{m}{n}+\frac{1}{n}=1.
\end{equation}

Thus, in 3D ($n=3$) homogeneous media, the velocity is constant (as are the density and elastic constants), so $p=0$ and, hence, $m=2$, as first shown by Debye. On the other hand, for the 1D chains considered above $n=1$ so $m=-p$, as found for both transverse and longitudinal vibrations, while both are 0 for the longitudinal ones. There is no apparent reason why $n$ should be restricted to integers.

In fact, both $p$ and $m$ are determined not only by the dimension $n$, but also by the dominant force governing the mechanics of the structure. The latter translates into the power index, $s$, in the relation $k\propto\omega^{s}$  for low frequencies, as in Eq. 4 (bending force) and 6 (stretching force), which cover the more usual cases for common molecules. In these cases, $s=0.5$ and 1, respectively. An immediate consequence is 
\begin{equation}
p=1-s,\,\, \mathrm{so} \,\, m=ns-1.
\end{equation}
The fundamental, independent parameters of the problem are $n$ and $s$. While the DOM depends on both, the velocity power index does not depend on $n$, which may take any value between 1 and 3. As a consequence, while $m$ can have a wide range of values, $p$ is essentially constrained to two values, 0.5 and 0, corresponding to transverse and longitudinal vibrations respectively.

Non-integer values of $n$ occur when, for instance, the structure consists in chains of catacondensed single carbon rings; $n$ then lies between 1 and 2, because, strictly, the structure is neither 1D nor 2D. In such cases, the constraint on $p$ helps lifting the ambiguity in $n$.

Some of these results are illustrated next.

\section{Minimum and maximum phonon frequencies}

\begin{figure}
\resizebox{\hsize}{!}{\includegraphics{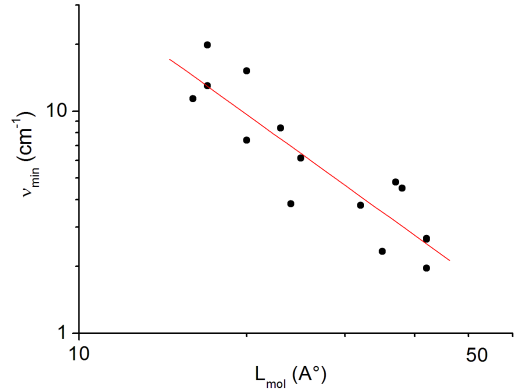}}
\caption[]{Dots: minimum wavenumber as a function of the largest molecular dimension. Line: power law fit, $\nu=2148/L_{\mathrm{mol}}^{1.8\pm 0.25}\,$ cm$^{-1}=6.4\,10^{4}/L_{\mathrm{mol}}^{1.8\pm 0.25}\,$GHz. The linear carbon chains which served as a basis for the theoretical analysis (Sec. 3 and 4) were not included in this graph as they do not seem to be abundant in space. However, they too fall well within the upper and lower limits of the power-law index in Eq. 18.}
\label{Fig:Lmolnu15}
\end{figure}

The minimum (or fundamental) phonon frequency, $\nu_{\mathrm{\mathrm{min}}}\equiv\nu(1)$, is that for which    $k=\pi\,f/L_{\mathrm{mol}}$, where $f$ is the correction factor for boundary effects (Sec. 2). Now, if $k\propto\omega^{s}$ (Sec. 3), then it follows that $V(\omega)=\omega/sk(\omega)$. Therefore,

\begin{equation}
 \nu(1)=\frac{sfV(1)}{2\mathrm{c}L_{\mathrm{mol}}}.
\end{equation}

 The corresponding mode is always that of a transverse phonon, because the stress governing longitudinal waves is stronger than that which governs transverse waves, making their velocity higher. If the velocity were independent of frequency ($s=1$), $\nu_{\mathrm{\mathrm{\mathrm{min}}}}$ would simply scale like the inverse of the molecular size, but this condition is far from common, except for 3D structures; as indicated above, for TA modes in 1D and 2D, the common value of $s$ is 0.5. 

In order to study relation 17 for more complex molecules than the linear ones in Sec. 2, 15  molecules were built, with backbones made up of 5- and 6-membered carbon rings, 16 to 42   \AA{\ } in major dimension (see Appendix). Figure \ref{Fig:Lmolnu15} collects the minimum wave numbers delivered by their normal mode analysis, as a function of molecular size. Although the structures are quite different and most of them are not plane, but bent and warped, a clear trend is apparent. The linear fit shown is
\begin{equation}
 \nu=2148\,L_{\mathrm{mol}}^{-1.8\pm0.25}
\end{equation}
 in cm$^{-1}$ and \AA{\ }, which is good to a factor 2. For $\nu$ in GHz, the pre-factor would be $6.4\,10^{4}$. 

\begin{figure}
\resizebox{\hsize}{!}{\includegraphics{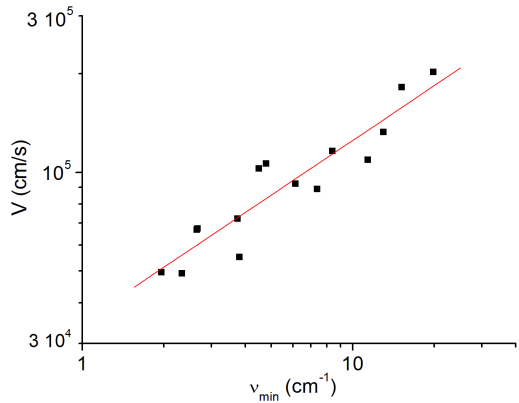}}
\caption[]{Dots: Transverse acoustic sound velocity derived from eq. 17, for the lowest phonon frequencies of the 15 carbon-based molecules of Fig. \ref{Fig:Lmolnu15} . Line: power law fit, $V=3.5\,10^{4}\,\nu^{0.55\pm 0.06}\,$ cm/s.}
\label{Fig:nuV15}
\end{figure}

On the other hand, inserting the same data ($L_{\mathrm{mol}},\,\nu_{\mathrm{\mathrm{\mathrm{min}}}}$) directly into Eq. 17, and plotting $V_{\mathrm{\mathrm{\mathrm{min}}}}$ against $\nu_{\mathrm{\mathrm{\mathrm{min}}}}$, one obtains Fig. \ref{Fig:nuV15}. Superimposed upon the dots is a linear fit to this plot: 
\begin{equation}
V_{\mathrm{\mathrm{min}}}=3.5\,10^{4}\,\nu_{\mathrm{\mathrm{\mathrm{min}}}}^{0.55\pm0.06}\,,
\end{equation}
whose exponent is close to the expected one, in agreement with the finding in Sec. 3 that $p=1-s$ is constrained to 0.5 for transverse waves. The ``average'' velocity is close to the velocity found for linear molecules, indicating that the dominant force in both 1D and 2D is that which resists bending, and that boundary effects are limited. Application of Eq. 12 to individual molecules confirms this result, as illustrated by examples in the following Section.

If $p=0.5$ is assumed to be the exact value, then Eq. 17 leads to $\nu_{\mathrm{\mathrm{\mathrm{min}}}}\propto L_{\mathrm{mol}}^{-2}$, with which Eq. 18 agrees within error bars. A number of causes may contribute to the dispersion of data points in Fig. \ref{Fig:Lmolnu15} and \ref {Fig:nuV15}: a) $L_{\mathrm{mol}}$ is not clearly defined  \emph{a priori}, unless the molecule is definitely a straight line or a plane surface; b) $\nu_{\mathrm{\mathrm{\mathrm{min}}}}$ is uncertain because of larger code errors at low frequencies and of mode coupling with fingerprint or longitudinal vibrations; c) boundary effects and uncertainty on the correction factor $f$; d) the velocity at a given frequency depends on mass (through $m$), density (through $a$) and bonding force strength (through $\epsilon$ or $\beta$), all of which depend on the particular structure.

The lack of computing power of presently available chemical modeling codes makes it difficult to explore larger structures. However, a conservative estimate may be obtained using Eq. 18 or Fig. \ref{Fig:Lmolnu15}. Thus, in order to contribute absorption/emission around 100 GHz and beyond, a structure has to be larger than about 40 \AA{\ }.

For a periodic chain of N atoms, the maximum phonon frequency corresponds to a vibration wavelength about equal to the atomic periodicity, $a$. For a TA phonon on a carbon chain, this is of order 1000 cm$^{-1}$, as in Fig. \ref{Fig:transv15}, which may be extended to carbon structures in general. In between the minimum and maximum frequencies, there are about 3N modes.
 Any slight change in shape or composition of the structure entails a slight redistribution of the whole spectrum. If the line of sight of a spectrometer includes enough molecules of the same structural type, then the spectrum will take the appearance of a continuum from about 1000 cm$^{-1}$ downwards. As suggested by Papoular \cite{pap14a}, this might apply to space, where huge quantities of molecules form around stars, novae and supernovae in a haphazard way but in the same local environment. And, indeed,  when observing galaxies or planetary nebulae one obtains continuous emission spectra starting from about 1000 cm$^{-1}$ and extending down to 1 cm$^{-1}$ or less: see Smith et al. \cite{smi}, Compi\`egne et al. \cite{com}, Fischer at al. \cite{fis}, Zhang and Kwok \cite{zha}.
 
\section{Some particular atomic structures}

\begin{figure}
\resizebox{\hsize}{!}{\includegraphics{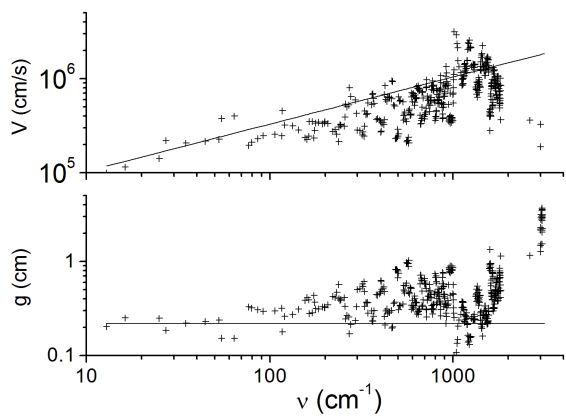}}
\caption[]{Benzene cluster 17$\times$17 \AA{\ }$^{2}$ in dimensions. Bottom: DOM, $g(\omega)$, delivered by the chemical model; line: power law fit to the phononic tail, $g=0.22$ cm. Top: same conventions for $V(\omega)$; line: power law fit to the phononic tail, $V=3.3\,10^{4}\,\nu^{0.5}$ cm/s.}
\label{Fig:cor1717nugV}
\end{figure}

 To illustrate finite 2D structures, three compact benzene clusters, 17$\times$17, 13$\times$17 and $7\times$34 \AA{\ }$^{2}$ in size, respectively, were built, optimized and analyzed for normal modes. The lower panel of Fig. \ref{Fig:cor1717nugV} displays the DOM, $g(\nu)$ of the $17\times$17  \AA{\ }$^{2}$
 structure, computed from its spectrum delivered by the simulation code. Here,  the spectral range of pure TA phonons is limited downwards by the structure finiteness and and upwards by boundary effects (edge hydrogens). Beyond that, the overlap of phonons of different types, as well as fingerprint modes, make interpretations hazardous. A straight line  $g=0.22$ cm is superimposed upon the DOM data, implying $m=0$, in agreement with Eq. 16 with $n=2$ and $s=0.5$.
 
\begin{figure}
\resizebox{\hsize}{!}{\includegraphics{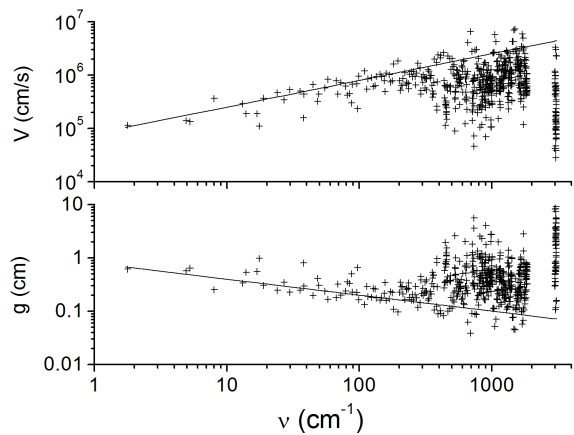}}
\caption[]{Chain of 11 anthracenes and phenanthrenes (126 C, 66 H). Bottom: DOM, $g(\omega)$, delivered by the chemical model; line: power law fit to the phononic tail, $g=0.8\,\nu^{-0.3}$ cm. Top: same conventions for $V(\omega)$; line: power law fit to the phononic tail, $V=8.1\,10^{4}\,\nu^{0.5}$ cm/s.}
\label{Fig:chphennugV}
\end{figure}

The upper panel displays the corresponding velocities, as deduced from $g(\nu)$ and eq. 12. The straight line represents a power law: $V=3.3\,10^{4}\,\nu^{0.5}$ cm/s, so $p=0.5$. Note that $m$ and $p$ exactly satisfy eq. 15 for $n=2$. This is not the case of the DOM and velocities of the other two benzene clusters, for their 2 dimensions are not exactly equal. If the average DOM of the 3 structures is considered, its low-frequency tail indicates $m=-0.1$. Inserting this value in Eq. 16, together with $s=0.5$, one obtains $n=1.8$, a measure of the excursion from the proper 2D case. This excursion increases, as expected, with the aspect ratio of the structure. Thus, for a chain of 11 anthracenes and phenanthrenes (126 C, 66 H), one finds $m=-0.3$, so, with $s=0.5$, we get $n=1.4$ ( Fig. \ref{Fig:chphennugV}).

The transverse vibrations of 2D molecules is reminiscent of plate and drum vibrations. However, it obeys, in fact, different mechanics because, whatever the thinness of a solid state sheet, the shear and compression will remain the dominant stresses.

\begin{figure}
\resizebox{\hsize}{!}{\includegraphics{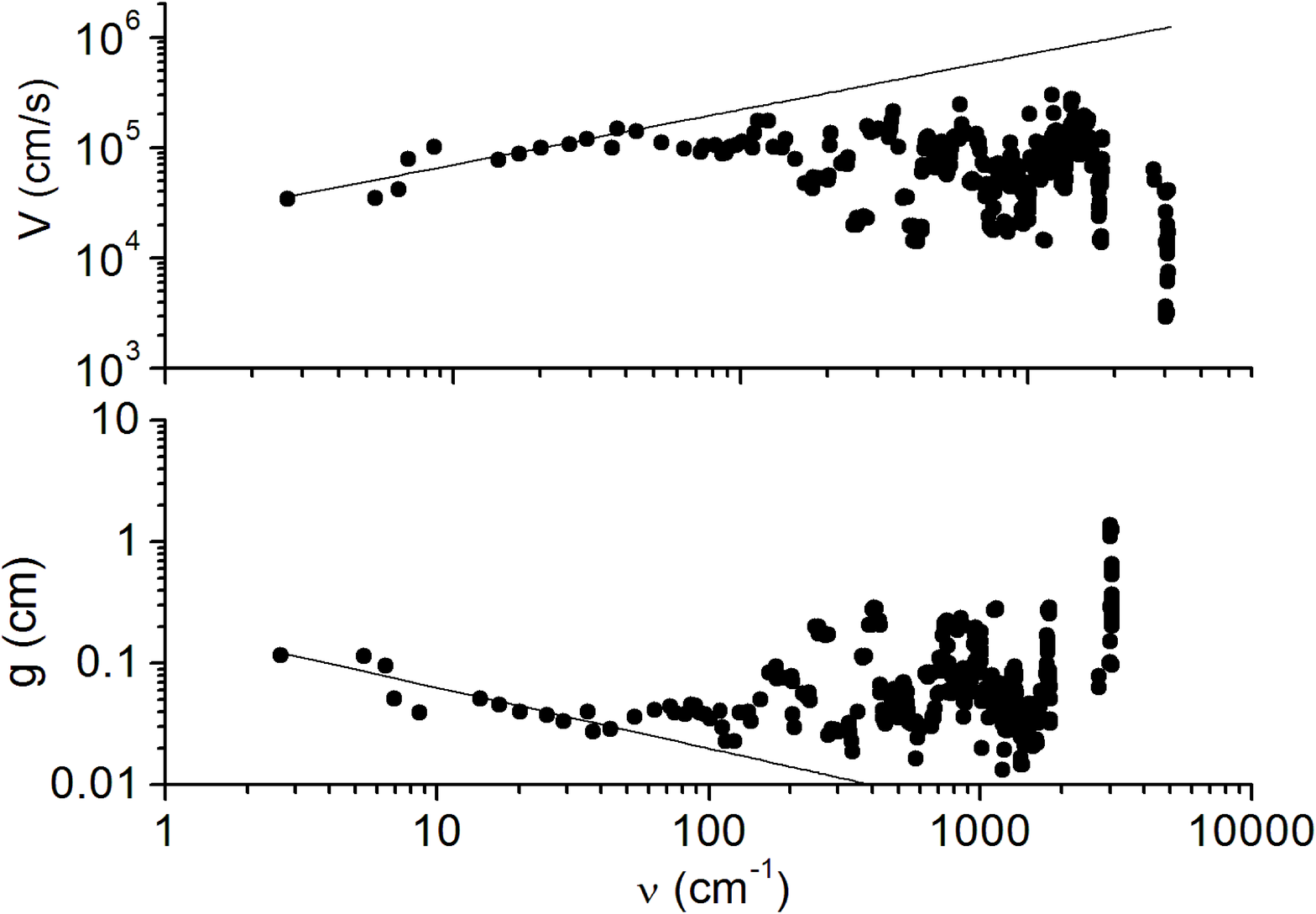}}
\caption[]{Chain of six ``trios'' connected in a row, one to another, by a single C-C bond; each trio is made of one 5-membered carbon ring squeezed between 2 6-membered rings. Bottom: DOM, $g(\nu)$, delivered by the chemical model (dots); line: power law fit to the phononic tail, $g=0.2\,\nu^{-0.5}$ cm. Top: same conventions for $V(\nu)$; line: power law fit to the phononic tail, $V=2.2\,10^{4}\,\nu^{0.5}$ cm/s.}
\label{Fig:triosnugV}
\end{figure}

As an example of unexpected findings, consider a chain of six ``trios'' connected in a row, one to another, by a single C-C bond. Each trio is made of one 5-membered carbon ring squeezed between 2 6-membered rings, the free vertex of the pentagon being occupied by a Sulfur atom. Each C-C connection is subject to bending \emph{and torsion}. Despite the planar character of each trio, relation 15 is only satisfied if the system is treated as one-dimensional, using the procedure of Sec. 3. Figure \ref{Fig:triosnugV} displays the results. In each graph, an average power-law line is superimposed upon the dots. The power-law indexes are -0.5 and 0.5, respectively, as for transverse phonons in linear chains, but the velocities are about a factor 0.7 lower than those of phonons in linear chains, at the same frequencies.

 \section{Extinction cross-section of phonons}
 In order to assess the spectroscopic detectability of these phononic vibrations, it is necessary to evaluate the intensity of the attending electromagnetic signal, if any. This depends, of course on the particulars of the molecules at hand. However, some general properties can be identified. The following study is restricted to carbon-based molecules which are 1- or 2-dimensional in the broad sense (i.e. not necessarily straight or plane, which is the case of most molecules in space).
 
 After optimization of the structure, the modeling software that is used here performs a Normal Mode analysis which delivers all mode frequencies with the corresponding integrated absorption intensities, $I(\nu)$ in km/mole (1 mole=6.024$\,10^{23}$ molecules)
. This is related to the absorptivity, $\alpha$, of a gas of identical molecules by
 \begin{equation}
 \alpha(\nu)=10^{2}\,I(\nu)\frac{\mathrm{C}}{\Delta\nu},
 \end{equation}
 where $\alpha$ is in cm$^{-1}$, C is the molecular density (mole/l) and $\Delta\nu$, the band width of the band at $\nu$, in cm$^{-1}$. For a large assembly of molecules of the same family but slightly different from one another, so as to produce a quasi-continuous vibrational spectrum, it can be shown (Papoular 2014b) that 
   \begin{equation}
 \alpha(\nu)=10^{2}\,<I(\nu)>g(\nu)\mathrm{C}.
 \end{equation}
 For astrophysical applications, it is often more convenient to consider the extinction cross-section per H atom, $\sigma$, which is related to $\alpha$ by
  \begin{equation}
\sigma(\nu)=f\Lambda\frac{\alpha(\nu)}{N_{\mathrm{C}}n_{\mathrm{dust}}} =\frac{f\Lambda}{6\,10^{18}}\frac{<I(\nu)> g(\nu)}{N_{\mathrm{C}}}, 
   \end{equation}      
 where $\sigma$ is in cm$^{2}$, $f$ is the fraction of carbon atoms locked in dust, $\Lambda$ is the cosmic abundance of carbon relative to hydrogen, 
 $ N_{\mathrm{C}}$ is the average number of C atoms in one dust particle and $n_{d}$ is the density of dust particle in space (cm$^{-3}$).

\begin{figure}
\resizebox{\hsize}{!}{\includegraphics{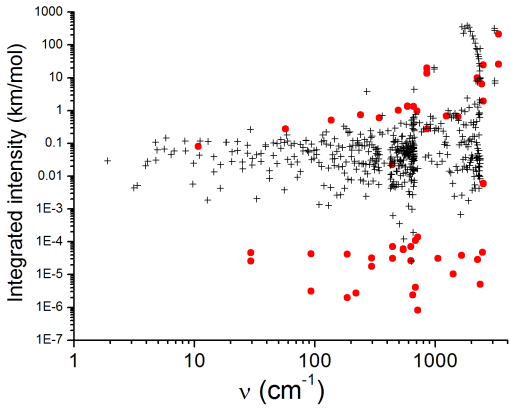}}
\caption[]{The integrated absorption intensity for the transverse phonons in linear carbon chains of 20 (red dots) and 160 atoms (black +).}
\label{Fig:nuIlin}
\end{figure}

 We now set out to estimate $\sigma$ for wavelengths beyond about 1000 $\mu$m. Consider, first, the intensities, $I$. Figure \ref{Fig:nuIlin} displays $I(\nu)$ for the shortest carbon chain studied in Fig.\ref{Fig:transv15} ( with CH terminations), and for the longest chain in the present study: 160 C atoms (terminated by a couple of H atoms at each end). 

 Two categories of transverse phonons can be distinguished: a) those with $I$ between about 0.01 and 0.1 km/mol; b) the ``suppressed'' ones, with $I$ one or several order of magnitude weaker. The distinction and the gap in $I$ are particularly clear for the shortest chain. A careful study of this case shows that the strongest (respectively the weakest) intensities are associated with odd (respectively, even) numbers of half-wavelengths along the chain, corresponding to constructive (respectively, destructive) combinations of atomic excursions. This effect weakens as the chain length increases, progressively reducing the range of intensities from below.

A third phonon category, the ``enhanced'' phonons, is discernible in the figure, as the wave number increases, but is more obvious when the molecular geometry favors the edges relative to the backbone or when the edges involve more infrared-active functional groups like OH for instance. In all these instances, the edge atoms dominate the molecular polarizability.

\begin{figure}
\resizebox{\hsize}{!}{\includegraphics{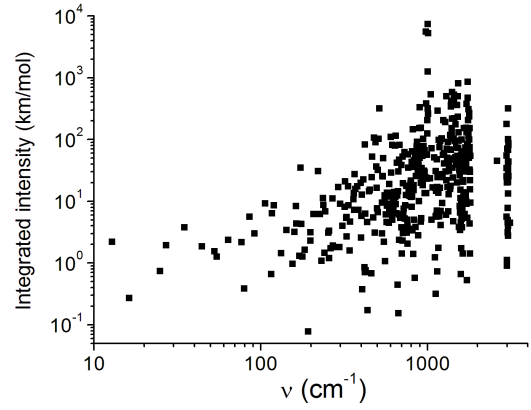}}
\caption[]{The integrated absorption intensities for  a benzene cluster 17$\times$17 \AA{\ }$^{2}$ in dimensions.}
\label{Fig:cor1717nuE}
\end{figure}

Similar trends are observed in Fig.\ref{Fig:cor1717nuE}, obtained from the modeling of the 17$\times$17 benzene cluster of Fig.\ref{Fig:cor1717nugV}. In this case, however, the absorption intensities of the low-frequency phonons are an order of magnitude higher and more and more high intensities occur as the frequency increases. Both effects are due to the increased number of atoms and to the presence of a larger relative number of edge hydrogen atoms.

From the consideration of these, and the other structures modeled in this work, and in view of the computational difficulty of increasing sizes, it may tentatively be concluded that the intensities of the lowest-frequency (transverse) phonons, for carbon based molecules in general, tend to flock around a constant value as the frequency decreases. This value lies between 0.01 and 0.1 km/mol for molecules with $n=1$, and 0.1 to 1 for structures with $n=2$.

As for the density of modes, $g(\nu)$, which also appears in Eq. 22, it can be derived using the results of Sec. 3 and 4. From Eq. 12
\begin{equation}
g(\omega)=\frac{d\omega}{dk}=\frac{nL}{\pi}\frac{G^{1-1/n}}{V(\omega)},
\end{equation}
and, from Eq. 13,
\begin{equation}
g(\omega)=g(\omega_{1})(\frac{\omega}{\omega_{1}})^{m}=\frac{nL}{\pi\,V(\omega_{1})}(\frac{\omega}{\omega_{1}})^{m},
\end{equation}
where $\omega_{1}$ is the lowest phonon angular frequency. Now, in the most common case of $n=2$, then $m=0$ and $g$ does not depend on frequency. From Sec. 4, we have, approximately,
\begin{equation}
\nu_{1}=\frac{2148}{L^{2}}\,\,\mathrm{and}\,\,V(\nu_{1})=3.5\,10^{4}\,\nu_{1}^{0.5}=\frac{1.62\,10^{6}}{L},              
\end{equation}
so whith $g$ in cm units and $L$ in \AA{\ },
\begin{equation}
g(\nu)=g(\nu_{1})=7.4\,10^{4}\,L^{2},
\end{equation}
in which, we note in passing, the particular case of Fig. 6 fits fairly well, as it should. Equation 22 also features $N_{C}$, which is found empirically to be $(L/2)^{2}$, so
\begin{equation}
\frac{g(\nu)}{N_{\mathrm{C}}}=3\,10^{-3}\,\,\mathrm{cm}.
\end{equation}
Inserting this result into Eq. 22, together with $\Lambda=1/3000$, and $<I>=0.1$ (see comments above in this section), one finally gets
\begin{equation}
\sigma_{\mathrm{C}}\simeq1.7\,10^{-26}\,f\,\,\mathrm{cm^{2}/H atom}.
\end{equation}
If one takes $f=0.15$, as did Papoular \cite{pap14b} to model the high Galactic latitude diffuse emission, then $\sigma\simeq2.6\,10^{-27}$ cm$^{2}$/H atom, independent of frequency and slightly weaker than the value found in that work for the longest wavelength, $\lambda=1000\,\mu$m  This confirms the preliminary conclusion drawn in Papoular \cite{pap14a}, to the effect that the extinction cross-section of carbon-rich dust particles levels off beyond a few hundred micrometers.
 It must be recalled that this extinction estimate is only valid to as long a wavelength as allowed by the maximum size of the particles present in the dust (Eq. 18). Towards shorter wavelengths, extinction progressively increases as functional group modes set in with increasing density and intensity, and the extinction settles into a logarithmic power law with index $\beta$ between 1 and 2, depending on environment.
 
 \section{Emission spectra}

 Knowing the integrated intensity spectrum of a molecule, it becomes possible to compute its emission spectrum if the excitation process is also known. The simplest case is that of thermal equilibrium with the environment at a given temperature. The emission spectrum is then obtained by multiplying the Black Body spectrum with the absorption cross-section (Eq. 22). An example of this procedure is given below. However, for this to be valid, the excitation rate must be strong enough to reach statistical equilibrium between emission and excitation processes. For low intensities of ambient radiation (G=G$_{0}$), this implies that the absorption cross-section of the molecule is large enough (size larger than 0.1-1 $\mu$m). Otherwise, before the molecule radiatively relaxes, it has time to settle in a state of internal statistical equilibrium, wherein the deposited excitation energy is continuously flowing between vibrational modes, but the energy in each mode remains constant on average. Roughly speaking, its emission spectrum will cover the range between 20 $\mu$m and $\lambda_{max}$, the maximum  wavelength in the phonon spectrum: this range will often turn out to be much wider than that of a black or gray body. 

\begin{figure}
\resizebox{\hsize}{!}{\includegraphics{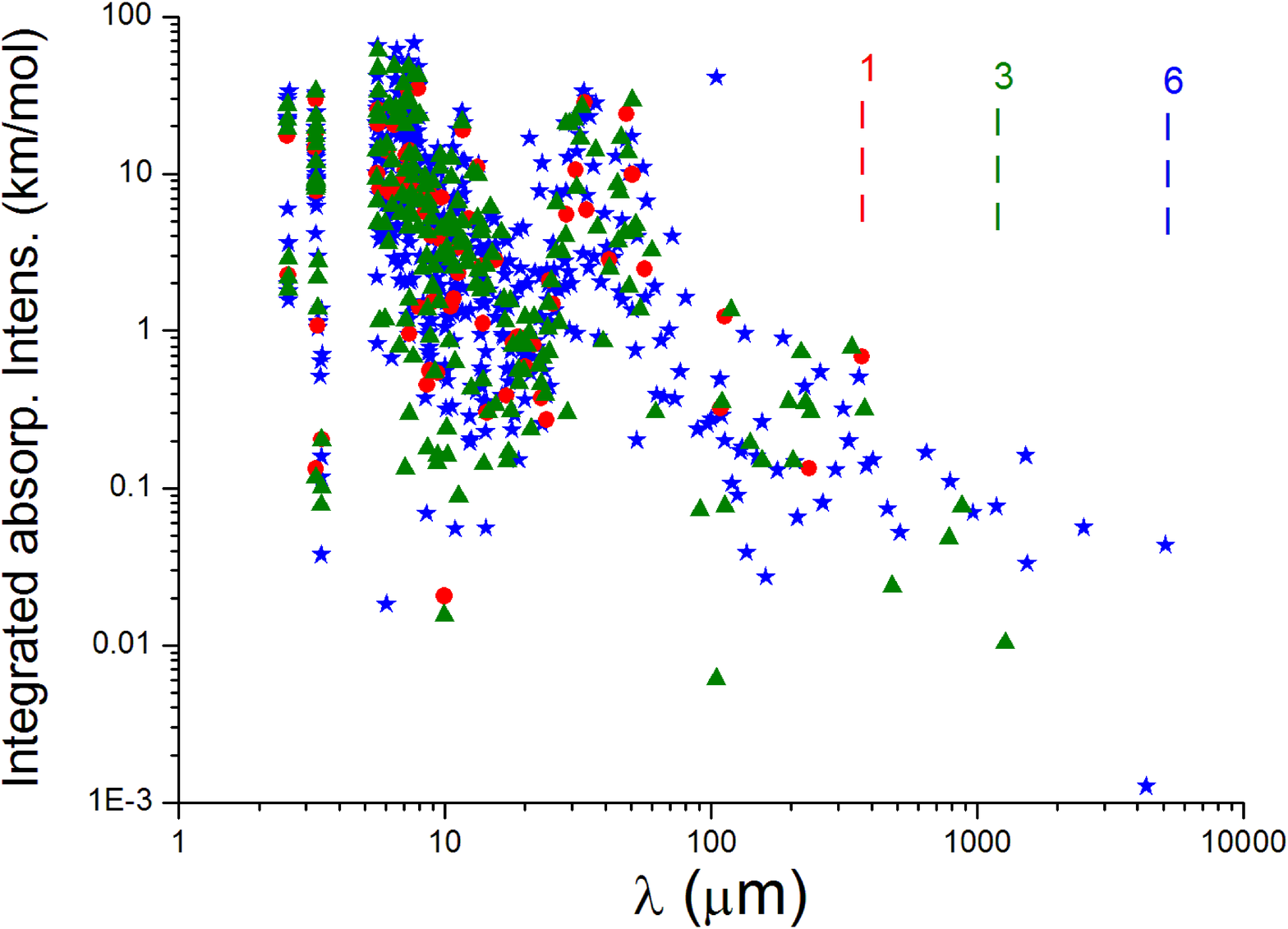}}
\caption[]{The integrated band intensities of the 3 selected molecules: 1 (red dots), 3 (olive triangles) and 6 ``trios'' (blue stars), respectively 9, 25,51 \AA{\ } in length. 
The abrupt end of each computed spectrum is indicated by vertical dashes. The maximum mode wavelength is 5097 $\mu$m.}
\label{Fig:kmmol}
\end{figure}

To illustrate this case, we select 3 organic molecules built up of 1, 3 and 6 ``trios'' (see the Appendix) arranged in a row so that their lengths turn out to be 9, 25 and  50  \AA{\ }, and their  $\lambda_{max}$ to be 369, 1275 and 5097, respectively, as shown in Fig. \ref{Fig:kmmol} which displays their integrated absorption spectra. They are meant to typify groups of dust particles of similar structure and sizes. The peak of integrated absorption which protrudes between 20 and 60 $\mu$m is a fingerprint of the OH groups attached to each trio, which are expected to be common in the ISM .

 \subsection{The computational procedure}

Now, it was shown (Papoular 2012) that the energy density ultimately radiated by an isolated molecule after it has been excited by radiation or collision that deposited an energy $E_{exc}$ in it, and after it has promptly relaxed into internal statistical equilibrium, is

 \begin{equation}
\mathrm{d}E^{*}(\nu)=E_{exc}\frac{e(\nu)\nu^3\mathrm{d}m(\nu)^{2}\mathrm{d}\nu}{\Sigma e(\nu)\nu^3\mathrm{d}m(\nu)^{2}\mathrm{d}\nu}\,,
\end{equation}
where $e(\nu)$ is the fraction of energy staying on average in mode $\nu$ and $m$ is the total electric dipole moment of the molecule.  It should be remembered that the latter (in the ground state of the molecule) is linked to the integrated band intensity, $I$ (km/mol), by
\begin{equation}
I(\nu)=2.6\nu m(\nu)^{2}\,,
\end{equation}
 stressing the intimate connection between emission and absorption. The modeling package used here provides for computing both distributions  $e(\nu)$ and $m(\nu)$. These are obtained in 3 successive runs of molecular dynamics:
 
 a) After optimizing the ground state of the selected molecule, the latter is given an energy $E_{exc}$ and left in vacuum to relax into its internal statistical equilibrium, which is reached within a few tens of picoseconds. 
 
 b)A second run is then launched, during which the fluctuations of the total kinetic energy in time are monitored. The Fourier transform of this delivers half the steady state kinetic energy distribution (e($\nu$)) among the modes, the other half being in the form of potential energy. This run must be at least twice as long as the period of the lowest frequency phonon.
 
 c)A third run is launched to monitor the temporal fluctuations of the total dipole moment. The Fourier transform of these delivers the $m(\nu)$ spectral distribution.
 
 \begin{figure}
\resizebox{\hsize}{!}{\includegraphics{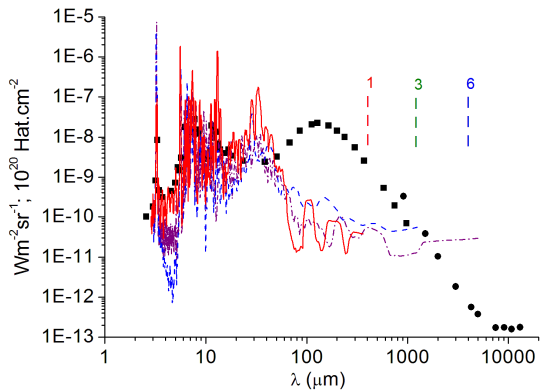}}
\caption[]{ Square dots: Diffuse High Galactic Latitude SED, adapted from Compi\`egne et al.  \cite{com}. Circular dots: example of measured AME, adapted from Abergel et al. \cite{abe14}. Spectra computed here for the three molecules made up of: 1 (red line), 3 (olive dotted line) and 6 (blue dash-dotted line) ``trios" in a row,  assuming G=G$_ {0}$ and $w=1$ (see text) in each case. Note the horizontal trend beyond the fingerprint range and the similarity, in relative values, of these emission spectra with the corresponding absorption spectra of Fig.\ref{Fig:kmmol}.The abrupt end of each computed spectrum is indicated by vertical dashes. It would take a molecule of 15 ``trios" in a row (390 atoms and 125 \AA{\ } in length) to cover the whole range of observed anomalous emission..}
\label{Fig:emissions}
\end{figure}

\subsection{Results}
Assume the rate of excitation events corresponds to G$_{0}$, or $10^{9}\,$cm$^{-2}$s$^{-1}$, and the total absorption c-s (cross-section) of all interstellar dust is $10^{-21}\,$ cm$^{2}$ per H atom (Spitzer 1978). The total rate of excitations is therefore $10^{-12}w\,$ s$^{-1}$, where $w$ is the fraction of dust cross-section contributed by a selected particle type. The emission power is obtained by multiplying this with $\mathrm{d}E^{*}(\nu]$ (Eq. 29); it is shown in Fig. \ref{Fig:emissions} for the 3 selected structures, assuming $w=1$ and $10^{20}$ H atoms per cm$^{2}$. The close similarity, in relative values, of these emission spectra with the absorption spectra (Fig. \ref{Fig:kmmol}) is remarkable and suggests that the latter are valuable guides in the study of emission. Another remarkable feature of the spectra is their flatness beyond the fingerprint region ($>$100 $\mu$m).

It has not proved possible to study longer molecules, as presently available modeling codes cannot handle particles larger than a few hundred atoms. However,  
the spectral extension to longer wavelengths clearly increases with the length of the molecule  as suggested by Sec. 4 and illustrated in Fig. \ref{Fig:kmmol}. Extrapolation of the present trend of $\lambda_{max}$ with molecular length indicates that merely increasing the latter to 130 \AA{\ } would suffice to cover the whole AME (Anomalous Microwave Emission; see Abergel et al. 2014, G\'enova-Santos et al. 2015) spectral range, to 10 GHz. 

 Figure \ref{Fig:emissions} also displays one version of the observed emission spectrum from the diffuse high latitude  Galactic medium (adapted from Compi\`egne et al. 2011 and Abergel et al. 2014). Comparison with the spectra computed here suggests that, with similar carbon-rich molecules, assuming G=G$_{0}$, only a small fraction, $w$, of the total dust is needed for their emission to reach the levels observed in the high latitude  Galactic medium, between 20 and 60$\mu$m, and beyond the 18-K gray-body spectrum (Compi\`egne 2012). Similar molecules include, of course, graphitic particles like PAHs and graphite.
 
  The evolution of the emission spectrum as a function of the molecular length illustrated in Fig. \ref{Fig:emissions} can be understood in the light of the phonon physics developed in Sec. 3 and 4, as shown in Appendix B. This also provides the grounds for extrapolation towards larger molecules and longer wavelengths.
  
\begin{figure}
\resizebox{\hsize}{!}{\includegraphics{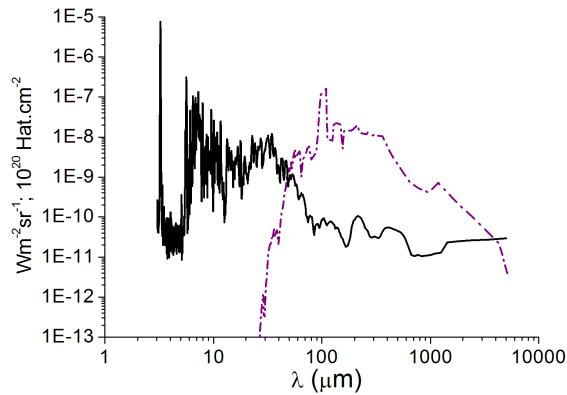}}
\caption[]{Emission of a 6-trio molecule under thermal equilibrium at 20 K (red dashes), or isolated in vacuum after thermalization at 20 K (black line).}
\label{Fig:cfemissns}
\end{figure}

  Figure \ref{Fig:cfemissns} compares, for the longest structure (six trios in a row), emissions at low and high (thermal equilibrium) rates of excitation, or, correspondingly, small and large particles. This highlights the difference in spectral extensions of emission for the same molecule.

 The present results are potentially of use in the on-going discussion of the carriers of AME and the degree of polarization of their emissions as compared  with the CMB. This problem recently gained traction in the wake of the measurement, by the BICEP2 telescope, of B-mode emission polarization in the spectral range of the cosmic microwave background (CMB), which could be a signature of early cosmic inflation (see Ade et al. 2014). The relevance, to this issue, of the molecules studied above depends first on their anisotropy. This may be illustrated by an example, viz. the 6-trios molecule of Fig. \ref{Fig:emissions}. Our chemical modeling package finds the 3 principal axes of inertia to be, respectively, 23036, 180748 and 192842 a.m.u. \AA{\ }$^{2}$, which also define the coordinates x,y and z. The mechanical anisotropy is obvious. The permanent dipole moment is 4.34 Debye, with components 1.98, -3.03 and 2.4.
  
  In order to assess the anisotropy of the vibrational emission, one follows in time each of the 3 components of the molecular dipole moment, then takes the Fourier transform, $I_{i}(\nu)$ (i:x,y,z) of its auto-correlation function. This delivers 3 curves similar and parallel to the blue curve in Fig. \ref{Fig:endmspec} of Appendix B. Now, after the molecule has been excited in a bath at 20 K (equivalent to a thermal spike), and left free in vacuum, it is found to rotate with an angular velocity of 2.3$\,10^{-3}$ rad/ps ($\sim$0.37 GHz). This is well within the range considered by Lazarian and Draine \cite{laz} both for their resonant relaxation and for Davis-Greenstein paramagnetic relaxation mechanisms. 

In the former mechanism, alignment in a  magnetic field occurs through rotation relaxation because of spin-lattice interaction. This requires that a quantum of rotation energy may be absorbed by the lattice, i.e. that resonance may occur between rotation and one of the lattice vibrations, for instance. For the 6-trios molecule considered here, this is not possible because the lowest vibrational frequency in the present example, is 60 GHz, much higher than the rotation frequency, 0.37 GHz. If  a bigger molecule is considered or some mechanism operates to give this molecule a higher rotation energy, then spin-lattice relaxation might become effective.

Anyhow, assume some alignment mechanism is at work and forces the molecular rotation vector to lie parallel with the Galactic magnetic field, which, in turn is supposed to be perpendicular to the sight line. In this picture, therefore,one component of the dipole moment lies along the magnetic field while the other two are randomly oriented in the plane perpendicular to this direction and containing the sight line.Three polarization fractions can then be deduced, according to which dipole moment component is parallel to the magnetic field:

\begin{equation}
 p_{i}(\nu)=\mid\frac{I_{j}(\nu)+I_{k}(\nu)-I_{i}(\nu)}{ I_{j}(\nu)+I_{k}(\nu)+I_{i}(\nu)}\mid; i\neq j\neq k\,.
  \end{equation}

Plotting the 3 quantities as a function of frequency, it is found that they oscillate between 0.4 to 0.95 at the phonon frequencies and nearly zero in between. Considering a family of similar molecules, differing only in length, and thus increasing the density of phononic states, one expects each polarization to retain high values throughout the spectrum. It must be stressed, however, that this is assuming perfect alignment and rigid structures. This whole exercise is only meant to show the potential of this type of molecules as polarized emitters in the AME spectral range.

\section{Conclusion}
Several neutral, carbon-based, hydrogen-terminated, energy-optimized molecules, made of chains and rings, were modeled by means of a semi-empirical algorithm. Their building blocks are simple enough that they may be synthesized in the laboratory and in space.

Their vibrational spectra were obtained by Normal Mode analysis. The focus was set upon those vibrations that extend over the whole structure (phonons), setting aside fingerprints due to small, inevitable, functional groups (contributing to the high frequency end of the spectrum). Transverse and longitudinal acoustic vibrations of the simplest structures could be separated by monitoring the atomic motions of each mode. The asymptotic behavior of their DOM and sound velocity toward low frequencies were found to obey power-laws in the frequency, with indexes $m$ and $p$, respectively. For the transverse vibrations, these indexes differ from the solid state values, 2 and 0. It was shown that this is so because the force governing the motion is that which resists bending of C-C bonds. The intensity of this force was confirmed independently. Because this force is weaker than the stretching force of C-C bonds, the lowest frequencies in a phonon spectrum are always those of TA vibrations. Knowledge of the asymptotic TA phonon velocity, in particular, is essential in determining the lowest frequency to be expected in the spectrum of a given molecule.

In the case of more complex structures, it is hardly possible to separate phonons of different types. The only quantity readily available from modeling is the DOM. Section 3 gives a general formula from which the sound velocity can be derived, given the DOM.

 Application of this formula requires the knowledge of the dimensionality, $n$, of the structure. For common structures, warped and finite, this notion is not as straightforward as it appears to be: $n$ is not always an integer, 1 or 2, but rather lies in between. It was shown that the ambiguity can be lifted by using a general relation between $m$, $n$ and $p$.
 
 These results were illustrated by the study of several structures of different types (but not crystalline). It was found that the TA sound velocity always scales like $\nu^{0.5}$, so the lowest frequency in a spectrum scales like $L^{-2}$, where $L_{\mathrm{mol}}$ is the longest straight dimension of the structure. The extinction cross-section of carbon-rich particles was shown to level off progressively beyond about 100 $\mu$m ($<$3000 GHz), and was estimated at $3\,10^{-27}$ cm$^{2}$/H atom, in order of magnitude, in this spectral range. 
 
 The results obtained above can readily be used to deduce emission spectra, assuming the ambient radiation density is high enough (or the dust particle large enough) that it reaches thermal equilibrium before re-radiating the energy gained. In the opposite case, the emission spectrum can be deduced from the monitoring of the temporal fluctuations of the total electric dipole moment after the molecule has been excited . From spectra computed in this way for molecules up to 50 \AA{\ } in length, it is concluded that reasonable amounts of C atoms locked in similar structures are sufficient to emit detectable radiation as far as 4000 $\mu$m at least. 
 Presently available modeling codes can hardly handle larger particles, but the theoretical analysis indicates that molecules 120 \AA{\ } in length are needed to cover the whole spectral range thought to be dominated by dust emission, before free-free bremsstrahlung and synchrotron emission prevail.

The primary application of this work is to the extension of the interstellar extinction curve and the variation of its slope. The relation between molecular size and lowest phonon frequency should help define the upper size limit of dust particles and perhaps their size distribution as well. The present results also shed new light on the vibrations of molecular structures that are not commonly studied. It was also shown that hydroxyl-bearing C-rich molecules can provide the emission observed between 20 and 60 $\mu$m, which is usually assigned to previously undefined very small grains (VSG; see D\'esert et al. 1990). 

One could also envision, in the future, an application to the elucidation of the sources and properties of AME. The challenge is to separate the contributions of CMB and foreground dust to the measured B-mode polarization (see Adam et al. 2014). Another challenge is to determine the cause(s) of emission polarization in the spectral range of interest. The present work indicates that large carbon-rich molecules may contribute significant polarized phonon emission in that range. Because of the intricacy and intertwining of CMB and dust emissions, an accurate quantitative knowledge of dust polarization is required. This is certainly an interesting subject of further investigation.

 \section{Appendix A}
 Table 1 provides the atomic composition, and the values of $L_{\mathrm{mol}}$ and 
  $\nu_{\mathrm{\mathrm{\mathrm{min}}}}$ of the molecules used in building Fig.\ref{Fig:Lmolnu15} and \ref{Fig:nuV15}, in order of increasing $\nu_{\mathrm{\mathrm{\mathrm{min}}}}$. 
  
  Here is a brief description of their structures:

1) 6 ``trios'' connected by single C-C bonds; each trio is made of a 5-membered ring squeezed between 2 6-membered ring; each has 1 edge hydroxyl and one -CH2OH;

2) simulation of a macromolecule of kerogene;

3) same as (1) but O atoms excluded;

4) same as (1), except: no hydroxyls, only two -CH2OH;

5)5 ``trios'' connected by single C-C bonds; each trio is made of a 5-membered ring squeezed between 2 6-membered ring; each has 1 edge hydroxyl and one -CH2OH;

6) 4 ``trios'' connected by single C-C bonds; each trio is made of a 5-membered ring squeezed between 2 6-membered ring; each has 1 edge hydroxyl and one -CH2OH;

7) linear chain of 15 benzenic rings;

8) linear chain of 4 coronenes; 

9) 1 pyrene, 1 trio, 1 benzenic ring; 1 pentagon;

10) 3 trios, 1 edge CH3;

11) 3 trios

12) hydrogenated amorphous carbon (HAC);

13) square cluster of benzenic rings (17$\times$17 \AA{\ }$^{2}$);

14) 2 coronenes, 1 edge -CH2OH;

15) cluster of benzenic rings (13$\times$17  \AA{\ }$^{2}$).

\begin{table}[ht]
\begin{flushleft}
\caption[]{The structures represented in Fig. \ref{Fig:Lmolnu15} and \ref{Fig:nuV15}}
\begin{tabular}{llllllll}
\hline
No & N$_{at}$ & C & H & O & S & $L_{\mathrm{mol}}$ & $\nu_{\mathrm{\mathrm{\mathrm{min}}}}$  \\ 
\hline
1 & 146 & 78 & 50 & 12 & 6 & 42 & 1.96 \\
\hline
2 & 493 & 254 & 185 & 40 & 9 & 35 & 2.34 \\
\hline
3 & 119 & 73 & 40 & 0 & 6 & 42 & 2.65 \\ 
\hline
4 & 130 & 76 & 46 & 2 & 6 & 42 & 2.67 \\
\hline
5 & 122 & 65 & 42 & 10 & 5 & 32 & 3.76 \\
\hline
6 & 98 & 52 & 34 & 8 & 4 & 48 & 1.27 \\
\hline
7 & 96 & 62 & 34 & 0 & 0 & 38 & 4.5 \\
\hline
8 & 142 & 103 & 38 & 1 & 0 &37 & 4.79 \\
\hline
9 & 101 & 51 & 39 & 8 & 2 & 25  & 6.15\\
\hline
10 & 62 & 37 & 22 & 0 & 3 & 20 & 7.4 \\
\hline
11 & 59 & 10 & 22 & 0 & 0 & 23 & 8.4 \\
\hline
12 & 292 & 110 & 182 & 0 & 0 & 16 & 11.4 \\
\hline
13 & 152 & 120 & 32 & 0 & 0 & 17 & 13 \\
\hline
14 & 74 & 49 & 24 & 1 & 0 & 20 & 15.2 \\
\hline
15 & 114 & 90 & 24 & 0 & 0 & 17 & 19.8 \\
\hline
\end{tabular}
\begin{list}{}{}
\item The units of $L_{\mathrm{mol}}$ and 
$\nu_{\mathrm{\mathrm{\mathrm{min}}}}$ are 
\AA{\ } and cm$^{-1}$, respectively.
 \end{list}
\end{flushleft}
\end{table}

 \section{Appendix B}
We seek here to understand the physical origins of the slope and ordinate at the longer wavelengths in the spectra in Fig. \ref{Fig:emissions}. We also seek to predict the emission power of molecules that would be large enough to contribute to the observed signals at the longest wavelengths of the AME spectrum. Sections 3 and 4 are helpful for this purpose, and the same symbol conventions will be used. 

\begin{figure}
\resizebox{\hsize}{!}{\includegraphics{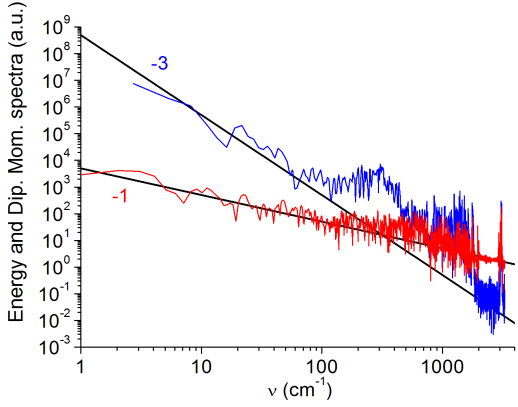}}
\caption[]{Spectra of energy (red line) and dipole moment squared (blue dashes) for the six-trios molecule, thermalized in a bath at 20 K, then isolated in vacuum and monitored; arbitrary units. The corresponding power indexes are shown.}
\label{Fig:endmspec}
\end{figure}

First, consider Fig. \ref{Fig:endmspec}, which displays some of the intermediate results that were used in drawing Fig.\ref{Fig:cfemissns}. It suggests that, in the range 1-100 $\mu$m, where the spectra are free of fingerprints features, the energy and electric dipole moment approximately follow power-law trends. Let the corresponding indexes be $E$ and $M$, which are found to be -1 and -3. $E$, which characterizes the energy distribution among modes in statistical equilibrium in vacuum, after excitation, is a result of the coupling between modes, and must be considered, here, as empirically given for it depends on the particular couplings and non-linearities of the molecule under consideration. $M$, however, is linked to $E$ by the relations obtained in Sec. 3.

In transverse mode No $i$ of a linear chain, the transverse atomic displacement is of the form $y_{i}=Y_{i}\mathrm{sin}(k_{i}ra)$, where $r$ is the rank of an atom along the chain. Then, from Eq. 1, the energy carried by the mode is 

\begin{equation}
E_{i}\propto Y^{2}_{i}\mathrm{sin}^{4}(k_{i}a/2)\Sigma_{r}\mathrm{sin}^{2}(k_{i}ra)\\
         \propto \sim LY^{2}_{i}(k_{i}a/2)^{4}\,,
\end{equation}
where L is the chain length and the sinus was approximated by its argument, as we are dealing with long wavelengths. Based on Sec. 3, $k_{i}\propto (i/L)$; from Sec. 4,
$i\propto (L^{2}\nu_{i})^{1/2}$. Now, the element of dipole moment squared, d$m^{2}_{i}$,  scales like $Y^{2}_{i}$. Figure \ref{Fig:endmspec} therefore suggests that we can put $Y^{2}_{i}=A\nu^{M}_{i}$. Hence,

\begin{equation}
E_{i}\propto L\frac{i^{4}}{L^{4}}A\nu^{M}_{i}\propto AL\nu^{2+M}_{i}\,.
 \end{equation}
From Fig.\ref{Fig:endmspec}, this scales like $\nu^{E}$, so $E=2+M$, which is satisfied by the indexes obtained empirically in that figure. 

Finally, Eq. 29 gives
 \begin{equation}
\nu\mathrm{d}E^{*}(\nu)\propto E_{exc}\nu\nu^{E}\nu^{3}LA\nu^{M}\propto E_{exc}\,,
 \end{equation}
independent of the frequency. That the emission spectrum is flat beyond the fingerprint region is very nearly confirmed, on average, by the modeling curves of Fig. \ref{Fig:emissions}. These various instances of relatively good agreement between analytical and modeling predictions provide grounds for extrapolating the present results to larger structures that cannot be handled by present modeling codes.


\begin{thebibliography}{}
 \bibitem[2011]{abe11}Abergel et al. 2011, Astronomy and Astrophysics 536, A21
\bibitem[2014]{abe14}Abergel et al. 2014, Astronomy and Astrophysics 566, A55
\bibitem[2014]{ada14}Adam et al. 2014,  arXiv 1409.5738
\bibitem[2011]{ade11}Ade P.A.R. et al. 2011, Astronomy and Astrophysics 536, A17
\bibitem[2014]{ade14}Ade P.A.R. et al. 2014, Astronomy and Astrophysics 565, 103
\bibitem[2014]{ade14b}Ade P.A.R. et al. 2014,  PRL 112, 241101
\bibitem[1990]{col}Colthup N., Daly L., Wiberley S. 1990, Introduction to IR and Raman Spectroscopy; Acad. Pr.: Boston
\bibitem[2011]{com}Compi\`egne M., Verstraete L., Jones A. et al. 2011,  A\&A 525, A103
\bibitem[1990]{des}D\'esert F.-X., Boulanger F. and Puget J.-L. 1990 A\&A 237, 215
\bibitem[1998]{dra}Draine B. and Lazarian A. 1998, ApJL 494, L19
\bibitem[1999]{fin}Finkbeiner et al. 1999, ApJ 524, 867 
\bibitem[2014]{fis}Fischer J. et al. 2014, arXiv 1409.2521
\bibitem[2015]{gen}G\'enova-Santos et al. 2015, arXiv 1501.0449
\bibitem[2007]{gei}Geim A. and Novoselov K.,2007, Nature Materials 6, 183
\bibitem[2012]{gho}Ghosh T. et al 2012, MNRAS 422, 3617
\bibitem[2005]{kit}Kittel C., 2005, Introduction to Solid State Physics; 8th ed.; Wiley: New York
\bibitem[2008]{kuz}Kuzmenko A., van Heumen E., Carbon F., van der Marel D. 2008, Phys. Rev. Lett. 100, 117401
\bibitem[2000]{laz}Lazarian A. and Draine B. 2000, ApJL 536, L15
\bibitem[2007]{moh}Mohr M, Maultzsch J., Dobardzic, E. et al. 2007, PR B 76, 035439
\bibitem[1972]{nic}Nicklow R., Wakabayashi N., Smith H.G.1972, PR B 5, 4951
\bibitem[2014a]{pap12}Papoular R. 2012, arXiv 1207.7217
\bibitem[2014a]{pap14a}Papoular R. 2014, MNRAS 434, 862
\bibitem[2014b]{pap14b}Papoular R. 2014, MNRAS 440, 2396
\bibitem[2007]{red}Reddy J.N. 2007 Theory and Analysis of Elastic Plates and Shells; CRC Press, Taylor and Francis
\bibitem[2007]{smi}Smith J. et al. 2007, ApJ 656, 770
\bibitem[1994]{spe}Speight J.  1994, Appl. Spectr. Reviews 1994, 29, 117
\bibitem[1978]{spi}Spitzer L. Jr. 1978, Physical processes in the interstellar medium, John Wiley and Sons, Inc. New York
\bibitem[1990]{ste}Stewart J. P., 1990, A semi-empirical molecular orbital program, Computer-aided molecular design 4:1-105
\bibitem[1953]{van}Van Hove L. 1953, Physical Review 89, 1189
\bibitem[1957]{wal}Wallis R. F. 1957, Phys. Rev. 105, 540
\bibitem[2011]{zha}Zhang Y., Kwok S. 2011, Nature 479, 80

\end{thebibliography}
\end{document}